\documentclass[aps,prl,twocolumn,superscriptaddress,showpacs,
               raggedbottom]{revtex4-1}
\usepackage{graphicx,amsmath,amssymb,bm,color}

\graphicspath{{figures/}}
\begin{document}

\title{The neutron polaron as a constraint on nuclear density functionals}

\author{M.\ M.\ Forbes}
\affiliation{Institute for Nuclear Theory, University of Washington,
Seattle, Washington 98195--1560, USA}
\affiliation{Department of Physics, University of Washington, Seattle,
Washington 98195--1560, USA}
\affiliation{Department of Physics and Astronomy, Washington State University, Pullman, Washington 99164-2814}
\author{A.\ Gezerlis}
\email[E-mail:~]{gezerlis@uoguelph.ca}
\affiliation{Department of Physics, University of Guelph, Guelph, 
Ontario N1G 2W1, Canada}
\affiliation{Institut f\"ur Kernphysik,
Technische Universit\"at Darmstadt, 64289 Darmstadt, Germany}
\affiliation{ExtreMe Matter Institute EMMI,
GSI Helmholtzzentrum f\"ur Schwerionenforschung GmbH, 64291 Darmstadt, Germany}
\author{K.\ Hebeler}
\affiliation{Institut f\"ur Kernphysik,
Technische Universit\"at Darmstadt, 64289 Darmstadt, Germany}
\affiliation{ExtreMe Matter Institute EMMI,
GSI Helmholtzzentrum f\"ur Schwerionenforschung GmbH, 64291 Darmstadt, Germany}
\affiliation{Department of Physics, The Ohio State University, 
Columbus, OH 43210, USA}
\author{T.\ Lesinski}
\affiliation{Institute for Nuclear Theory, University of Washington,
Seattle, Washington 98195--1560, USA}
\affiliation{Department of Physics, University of Washington, Seattle,
Washington 98195--1560, USA}
\affiliation{University of Warsaw, Institute of Theoretical Physics,
ul. Ho\.za 69, 00-681 Warsaw, Poland}
\affiliation{CEA Saclay, IRFU/Service de Physique Nucl\'eaire, F-91191 Gif-sur-Yvette, France}
\author{A.\ Schwenk}
\affiliation{ExtreMe Matter Institute EMMI,
GSI Helmholtzzentrum f\"ur Schwerionenforschung GmbH, 64291 Darmstadt, Germany}
\affiliation{Institut f\"ur Kernphysik,
Technische Universit\"at Darmstadt, 64289 Darmstadt, Germany}

\begin{abstract}
  We study the energy of an impurity (polaron) that interacts strongly
  in a sea of fermions when the effective range of the
  impurity-fermion interaction becomes important, thereby mapping the
  Fermi polaron of condensed matter physics and ultracold atoms to
  strongly interacting neutrons. We present Quantum Monte Carlo
  results for this neutron polaron, and compare these with effective
  field theory calculations that also include contributions beyond the
  effective range. We show that state-of-the-art nuclear density
  functionals vary substantially and generally underestimate the
  neutron polaron energy. Our results thus provide constraints for
  adjusting the time-odd components of nuclear density functionals to
  better characterize polarized systems.
\end{abstract}


\maketitle
\noindent
Energy-density functionals are the only method available to study
heavy nuclei and to globally describe the chart of
nuclides~\cite{Bender:2003,Erler}. Due to the need for a precise
description of low-energy observables, parameters of these functionals
are generally fit to properties of nuclei, including masses and
radii. This empirical construction can therefore also benefit from
additional input to constrain their properties in exotic, i.e.,
neutron-rich or spin-polarized systems. Examples of such pseudo-data
include calculations of neutron
matter~\cite{dEFT,Gezerlis1,Hebeler:2010,GCR,Lacour,Tews:2013,PPNPMunich}
and neutron drops~\cite{ndrops}. This approach has been successfully
used to shape new functionals (see, e.g.,
Refs.~\cite{Fayans,SLy,Gogny,Goriely,Bulgac,Fattoyev}). In this
work, we study the neutron polaron and use its energy as a
constraint on nuclear density functionals.

The polaron was first introduced in condensed matter physics, and has recently
been investigated in strongly interacting ultracold Fermi
gases~\cite{Chevyreview} -- a system that has many similarities with the physics
of low-density neutron matter (see, e.g., Refs.~\cite{dEFT, Gezerlis1}). The
Fermi polaron is an impurity interacting in a Fermi sea, realized in ultracold
atoms and neutron matter as a spin-down fermion in a sea of $N_\uparrow$ spin-up
fermions. The polaron energy $E_\text{pol} = E_{N_\uparrow+1} - E_{N_\uparrow}$
is defined as the energy difference between the system with the polaron added
and the $N_\uparrow$ (non-interacting) Fermi system. In the thermodynamic limit,
this is equivalent to the spin-down chemical potential in the limit of high
polarization, and therefore constrains the phase diagram of strongly interacting
Fermi systems as a function of spin imbalance~\cite{Chevy,BF,CR,Lobo,Pilati}.

For attractive interactions, $E_\text{pol} < 0$ measures the polaron
binding energy in the Fermi sea. In the unitary limit, where the S-wave
scattering length $|a| \to \infty$, the polaron energy is universal at
low densities and scales as $E_\text{pol} = \eta E_F$ where
$E_F = k_F^2/2m$ is the Fermi energy (with Fermi momentum
$k_F$) and $\eta < 0$ is a universal number~\cite{Chevy}.
Neutrons, whose scattering length is large ($a = -18.5 \, \text{fm}$), 
have low-density properties close to the unitary limit.

At unitarity, the polaron energy admits a variational bound that sums
one-particle--one-hole excitations, $\eta \leq -0.6066$~\cite{Chevy}, which is
remarkably close to a full many-body treatment~\cite{Combescot}, $\eta =
-0.6158$, and agrees with Quantum Monte Carlo (QMC)
calculations~\cite{Lobo,CR,Pilati,PS}.  These theoretical values are consistent
with experimental extractions of $\eta=-0.58(5)$~\cite{Shin:2008} and
$-0.64(7)$~\cite{Schirotzek:2009} from ultracold atoms across a Feshbach
resonance. The Fermi polaron continues to be an exciting area of
research~\cite{PolaronGrimm,PolaronReview}, with recent studies of the polaron
in two dimensions~\cite{2dpolaron,2dpolaron_exp} and of the P-wave
polaron~\cite{Pwavepolaron}.

Here, we generalize the polaron to strongly interacting neutrons,
where the effective range $r_e = 2.7 \, \text{fm}$ is important, and $k_F r_e
\sim 1$ is not small, as is relevant for nuclei. We calculate the polaron energy
using an effective field theory (EFT) for large~$a$ and large~$r_e$, and from
chiral EFT interactions that include contributions beyond the
effective range. We benchmark these approximations with QMC simulations, and
compare the resulting $E_\text{pol}$ to predictions of nuclear density
functionals.  Finally, we construct a functional that includes the polaron
energy as a constraint.

The Chevy Ansatz~\cite{Chevy} for the polaron energy can be
generalized to include a large effective range using a di-fermion EFT
(dEFT) where the fermions $\psi$ interact through a di-fermion field
$d$ and the energy dependence of the di-fermion propagator generates
the effective range~\cite{dEFT,KBS}. The lowest-order dEFT Lagrangian
density is given by
\begin{align}
  \mathcal{L} &= \psi^\dagger \biggl( i \partial_0 + \frac{\nabla^2}{2m}
  \biggr) \, \psi - d^\dagger \biggl( i \partial_0 + \frac{\nabla^2}{4m}
  - \Delta \biggr) \, d \nonumber \\
  &\quad - g \, \bigl( d^\dagger \psi\psi + d \, \psi^\dagger \psi^\dagger
  \bigr) \,,
  \label{dEFT}
\end{align}
where $\Delta$ and $g$ describe the propagation of the di-fermion field and its
coupling to two fermions, respectively. Matching these to the effective-range
expansion with a cutoff regularization (for large cutoffs $\Lambda$) gives
$\Delta/(m g^2) =1/(4 \pi a) - \Lambda/(2 \pi^2)$ and $(m/g)^2 = r_e/(8 \pi) -
1/(2 \pi^2 \Lambda)$.  Upon integrating out the di-fermion field, we obtain an
energy-dependent potential between the fermions $V(E) = g^2/(\Delta - E)$, where
$E$ is the energy in the center-of-mass system. Using $V(E)$ with the Chevy
wave function, $|\psi \rangle = \alpha_0 |\Omega \rangle + \sum_{{\bf p},{\bf
    h}} \alpha_{{\bf p},{\bf h}} |{\bf p}, {\bf h} \rangle$, of spin-up
one-particle--one-hole excitations $|{\bf p}, {\bf h} \rangle$ ($p > k_F$, $h
\leqslant k_F$) on top of a Fermi sea $|\Omega \rangle$, we find for the polaron
energy the self-consistent equation
\begin{align*}
  E_\text{pol} &= \int_0^{k_F} \frac{h^2 dh}{2 \pi^2 m} \Biggl[
  \frac{1}{4 \pi a} - \frac{r_e}{8 \pi} \biggl( m E_\text{pol} +
  \frac{h^2}{4} \biggr) - \frac{k_F}{2 \pi^2} \nonumber \\
  &- \int_{k_F}^\infty
  \frac{p^2 dp}{(2 \pi)^2} \biggl( \frac{1}{ph} \ln \biggl| \frac{p^2 - ph
    -m E_\text{pol}}{p^2 + ph -m E_\text{pol}} \biggr| + \frac{2}{p^2} \biggr)
  \Biggr]^{-1}.
\end{align*}
This is equivalent to the Dyson equation, $E_\text{pol} = \Sigma(
E_\text{pol})$, where $\Sigma$ is the self-energy of the spin-down
polaron with zero momentum in a Fermi sea of spin-up
particles~\cite{Chevyreview}. Since the density of spin-down particles
vanishes, diagrams involving intermediate spin-down hole states do not
exist in the polaron limit. The Chevy ansatz corresponds to
calculating $\Sigma$ at the $T$ matrix level, without particle-hole
corrections for the single-particle energies, so that at any time in
a diagram, there is only one spin-up particle-hole excitation from
the noninteracting $|{\bf p}, {\bf h} \rangle$ in the wave function.

\begin{figure}[t]
\begin{center}
\includegraphics[width=\columnwidth]{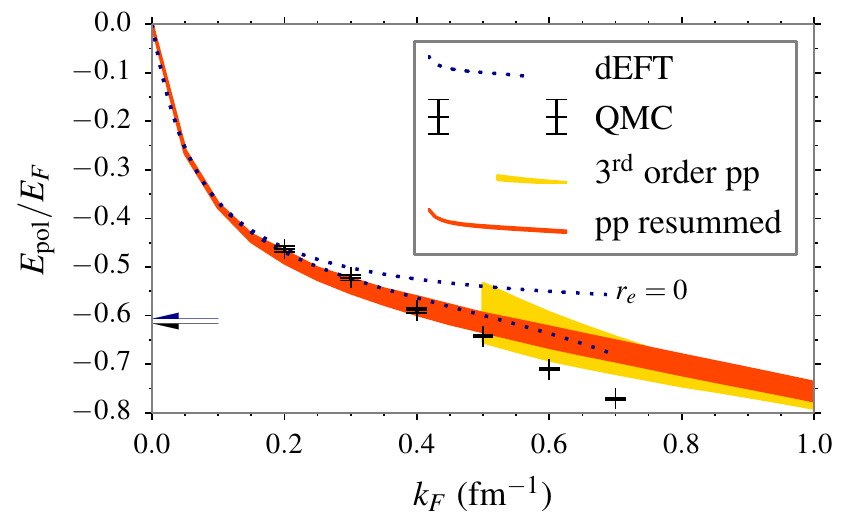}
\caption{(Color online) Neutron polaron energy $E_\text{pol}$ (in units
of the Fermi energy $E_F$) as a function of Fermi momentum
$k_F$. The arrows mark the values at unitarity $\eta 
= -0.607$ and $-0.616$ (see text). Results are shown from
the dEFT calculation with and without an effective range (the $r_e=0$
result approaches the unitarity value with increasing $k_F$) and
from QMC calculations with 33+1 particles and S-wave interactions. The
two bands are based on chiral NN interactions at N$^3$LO, including
also contributions beyond effective-range effects, at the level of
third-order particle-particle (pp) ladder contributions (yellow band;
3rd order pp) and resumming pp ladders (red band; pp resummed). The
width of the bands reflects the variation from using different
cutoffs, different single-particle energies, and for the 3rd order pp
band, the difference from second- to third-order ladders.\label{fig:pol}}
\end{center}
\end{figure}

The dEFT results for the neutron $E_\text{pol}$ as a function of $k_F$ are shown
in Fig.~\ref{fig:pol} with and without an effective range. The $r_e=0$ result
approaches the unitary value $\eta = -0.607$~\cite{Chevy} with increasing $k_F$,
so that $1/(k_F a) \to 0$. For positive $r_e$, the neutron polaron binding
increases, as observed in Fig.~\ref{fig:pol} for $k_F > 0.2 \, \text{fm}^{-1}$,
where $k_F r_e \approx 0.5$. Intuitively, the polaron interacts with more
particles within the range of the interaction. Our results are consistent with
Ref.~\cite{dipolar}, which studied the potential realization of a positive
effective range induced by resonant dipolar interactions in ultracold atoms,
and~\cite{Massignan:2012,Trefzger:2012}, which studied narrow resonances.
Conversely, for negative $r_e$, which is realized with a narrow Feshbach
resonance, the polaron binding weakens.

The dEFT makes two approximations: first, it neglects interaction
effects beyond the large $a$ and $r_e$; second, it is restricted to
contributions from one-particle--one-hole excitations. We address the
former with microscopic calculations using chiral EFT interactions to
next-to-next-to-next-to-leading order (N$^3$LO). Our calculations are
based on the $500 \, \text{MeV}$ N$^3$LO NN potential of
Ref.~\cite{EM} and include all partial waves with total angular
momentum $J \leqslant 6$. Three-nucleon forces are expected to be
small at the densities considered~\cite{Hebeler:2010,Tews:2013}. To
study how perturbative the many-body problem is, we also use the
renormalization group (RG)~\cite{PPNP} to evolve the NN potential to
low-momentum interactions $V_{\text{low} \, k}$ with cutoffs
$\Lambda=1.8-2.8
\,\text{fm}^{-1}$. References~\cite{Hebeler:2010,Tews:2013} show that
neutron matter is perturbative at nuclear densities for these
interactions, as was recently validated using QMC calculations with
chiral EFT interactions~\cite{QMCchiral}.

Figure~\ref{fig:pol} gives our results for the neutron polaron energy based on
unevolved and RG-evolved chiral NN interactions at N$^3$LO. The red band (pp
resummed) is obtained at the $T$ matrix level, resumming particle-particle (pp)
ladders. The width of the band reflects the variation from using the different
cutoffs and either free or Hartree-Fock single-particle energies. The pp
resummed results agree with perturbative calculations including up to
third-order pp ladder contributions (see Ref.~\cite{nucmatt} for details), shown
by the yellow band (3rd order pp). This latter band also includes an estimate
for the perturbative convergence (given by plus/minus the difference from
second- to third-order). Similarity RG-evolved interactions lead to analogous
results. The perturbative results are shown only for $k_F \geqslant 0.5 \,
\text{fm}^{-1}$, because the pp channel becomes nonperturbative at low densities
due to the large scattering length~\cite{PPNP}. The microscopic calculations
based on chiral NN interactions agree with the dEFT results, indicating that
contributions beyond effective-range effects are small at these densities.

To benchmark the polaron energy and address the previous restriction to
one-particle--one-hole excitations, we perform Green's Function Monte Carlo
(GFMC) calculations following Refs.~\cite{Gezerlis1,Gezerlis2,Gezerlis3} with
the following Hamiltonian:
\begin{equation}
  \mathcal{H} =
  - \frac{\hbar^2}{2m}\nabla_{1'}^{2}
  -\frac{\hbar^2}{2m}\sum_{i=1}^{N_{\uparrow}}\nabla_i^{2}
  + \sum_{i=1}^{N_{\uparrow}} V(r_{i1'}) \,,
\end{equation}
where $r_{i1'}$ is the distance between the $i$'th majority particle
and the impurity ($1'$).  We take the S-wave interaction to be either
the $^1$S$_0$ channel of the Argonne $v_{18}$ neutron-neutron
potential or a modified P\"{o}schl-Teller potential fit to the
scattering length and effective range of the $v_{18}$ potential. Both
results agree, demonstrating universality (independence of the
potential) over the densities studied with QMC ($0.2 \,
\textrm{fm}^{-1} \leqslant k_\textrm{F} \leqslant 0.7 \,
\textrm{fm}^{-1}$). Universality is further confirmed by our finding
that same-spin P-wave interactions are negligible.

The QMC algorithm evolves an initial state in imaginary time, finding
the lowest energy within the space of wave functions having the same
nodal structure as the initial state. Since this system has no
pairing, we start with a Slater determinant of plane waves for the
majority species (as in Ref.~\cite{Gezerlis3}), and a single plane
wave $e^{i \mathbf{k}\cdot\mathbf{r}_{1'}}$ for the impurity.  The
algorithm thus provides an upper bound for the energy $E_{N_\uparrow}$
of the system with $N_\uparrow + 1$ particles, and therefore to the
polaron energy $E_\textrm{pol} = E_{N_\uparrow + 1} - E_{N_\uparrow}$.
(Note that these energies are compared at fixed simulation volume $L^3
= 6\pi^2 N_{\uparrow}/k_F^3$, not at fixed total
density~\cite{Gezerlis4}.) Our results for $E_\textrm{pol}$ obtained
from $N_\uparrow = 33$ are presented in Fig.~\ref{fig:pol}.

By varying the momentum of the impurity, we can extract its effective mass by
fitting a line to $E_\textrm{pol}(k)$ vs\@. $k^2$ for momenta (1,2,3,4 times
$2\pi/L$) at each density.  We find the same increased effective mass $m^*/m =
1.04(3)$ as in the unitary limit~\cite{Lobo} for all densities considered,
consistent with microscopic calculations of the Fermi liquid parameters of
neutron matter (see, e.g., Ref.~\cite{RGnm}).

\begin{figure}[t]
\begin{center}
\includegraphics[width=\columnwidth]{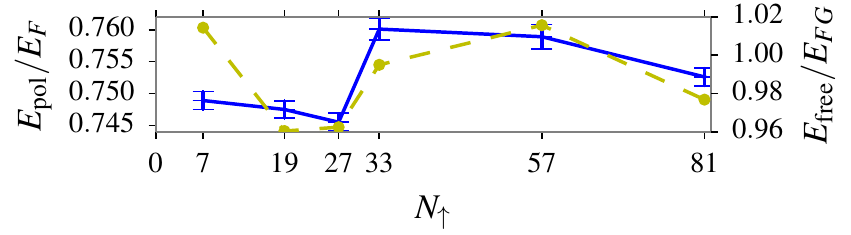}
\caption{(Color online) Dependence of the polaron energy $E_\text{pol}$
(solid line, left axis, in units of the Fermi energy $E_F$) on
the number of spin-up neutrons $N_{\uparrow}$ for the largest density
($k_F= 0.7 \, \text{fm}^{-1}$). This follows the dependence of
noninteracting neutrons $E_\text{free}$ (dashed line, right axis, in
units of the Fermi-gas energy $E_{FG}$) except for the smallest
system $N_{\uparrow}=7$, where the box is only 4 times larger than the
effective range.\label{fig:finite}}
\end{center}
\end{figure}

The QMC results agree with the dEFT and the bands in Fig.~\ref{fig:pol} for low
densities $k_F \lesssim 0.5 \, \text{fm}^{-1}$, but yield a lower energy as the
density increases. These differences could be due to nonperturbative
many--particle-hole effects neglected by the latter approaches. The QMC
calculations are performed in a box, but exhibit very small finite-size
effects. The dependence on particle number, shown for the largest density
($k_{F}= 0.7 \, \text{fm}^{-1}$) in Fig.~\ref{fig:finite}, follows the
noninteracting system, justifying the use of the 33+1 particle QMC results to
approximate the thermodynamic limit. This finding is consistent with
Ref.~\cite{Gezerlis1} for the spin-symmetric paired system.

Finally, to assess the impact of tensor and spin-orbit interactions,
we use Auxiliary-Field Diffusion Monte Carlo (AFDMC)~\cite{Schmidt:1999,%
Gandolfi} to calculate the polaron energy with the Argonne $v_8'$
potential~\cite{Wiringa:2002}. We have found that the GFMC algorithm leads to more
accurate results for large polarizations, but AFDMC has the advantage that
it can include tensor and spin-orbit interactions nonperturbatively.
Comparing the AFDMC [vs.\ GFMC] energies for 33+1 particles at two intermediate
densities we find $E_\text{pol}/E_F = -0.531 \pm 0.008$ [vs.\ $-0.522 \pm
0.006$] for $k_F = 0.3 \, \text{fm}^{-1}$ and $E_\text{pol}/E_F = -0.567 \pm
0.006$ [vs.\ $-0.589 \pm 0.005$] for $k_F = 0.4 \, \text{fm}^{-1}$. We have also
performed AFDMC simulations at these $k_F$ values using the Argonne $v_6'$ (no
spin-orbit) and $v_4'$ (plus no tensor) interactions, and find consistent
results indicating that tensor and spin-orbit interactions have a small effect
at these densities. This implies that the polaron lifetime is long, which is
consistent with expectations based on calculations of the small spin relaxation
rate $\Gamma_\sigma \ll E_\text{pol} \sim E_F$ in neutron matter at low
densities~\cite{Bacca}.

Having established the polaron energy in Fig.~\ref{fig:pol}, we study
the impact on nuclear density functionals. We consider the family of
Skyrme functionals, which have been used in global studies of
nuclei~\cite{Bender:2003}. The energy density of neutron matter
$\mathcal{E}$ is given by the parametrization~\cite{Perlinska:2004}
\begin{align}
  \mathcal{E} &= \frac{\hbar^2}{2m} \, \tau 
  + (C^\tau_0 + C^\tau_1) \, \rho \, \tau 
  + (C^{sT}_0 + C^{sT}_1) \, \mathbf{s}\cdot\mathbf{T} \nonumber \\ 
  &\quad+ (C^{\rho,0}_0 + C^{\rho,0}_1) \, \rho^2 
  + (C^{\rho,D}_0 + C^{\rho,D}_1) \, \rho^{2+\gamma} \nonumber\\
  &\quad+ (C^{s,0}_0 + C^{s,0}_1) \, \mathbf{s}^2 + (C^{s,D}_0 + C^{s,D}_1)
  \, \mathbf{s}^2\rho^\delta \,,
  \label{Skyrme}
\end{align}
with density $\rho = \rho_\uparrow + \rho_\downarrow$, spin density
$\mathbf{s} = \rho_\uparrow - \rho_\downarrow$, kinetic density $\tau
= \tau_\uparrow + \tau_\downarrow$, and spin kinetic density
$\mathbf{T} = \tau_\uparrow - \tau_\downarrow$. The various
functionals differ in the set of Skyrme parameters $C$, which
follow from fits to selected properties of nuclei and nuclear
matter. For neutron matter only the isoscalar plus isovector
($C_0+C_1$) combinations enter. We have allowed the usual
density-dependent $C^D$ terms to have different powers of the density
for the time-even ($\gamma$) and time-odd ($\delta$) parts.

\begin{figure}[t]
\begin{center}
\includegraphics[width=\columnwidth]{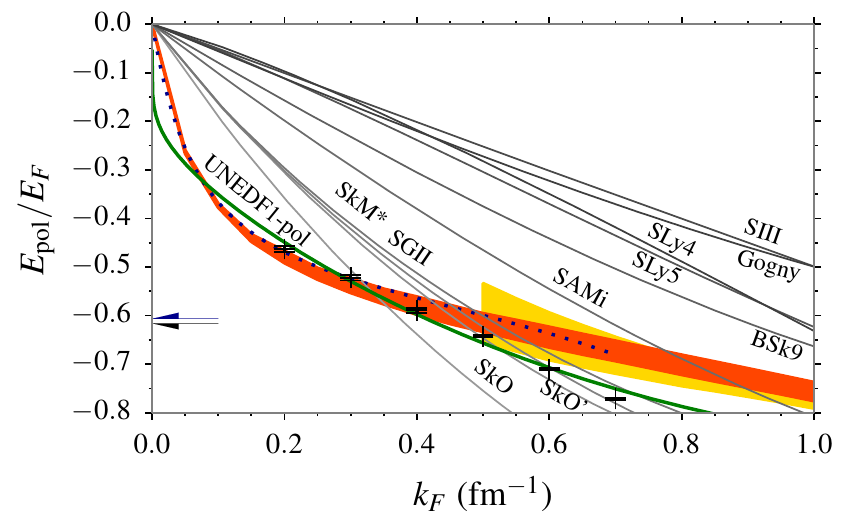}
\caption{(Color online) Comparison of the microscopic results for the
polaron energy of Fig.~\ref{fig:pol} with predictions of nuclear density
functionals~\cite{SIII,SGII,SkM,SLy,SkO,BSk9,SAMi,Gogny} (see text).
The thick solid (green) curve is the new UNEDF1-pol density functional,
which is based on Ref.~\cite{UNEDF1} plus a fit to the QMC $E_\text{pol}$
results.\label{fig:EDF}}
\end{center}
\end{figure}

The polaron energy follows from the energy density~(\ref{Skyrme})
by $E_\text{pol} = (\partial\mathcal{E}/\partial\rho_\downarrow)
\bigr|_{\rho_\downarrow=0}$.  Figure~\ref{fig:EDF} shows the
predictions for $E_\text{pol}$ of various state-of-the-art nuclear
density functionals: SIII~\cite{SIII}, SGII~\cite{SGII},
SkM$^*$~\cite{SkM}, SLy4 and SLy5~\cite{SLy}, SkO and
SkO$'$~\cite{SkO}, BSk9~\cite{BSk9}, SAMi~\cite{SAMi}, as well as the
Gogny D1N functional~\cite{Gogny}.  All functionals predict an
attractive polaron energy, but $E_\text{pol}$ varies greatly among the
different functionals and is generally underestimated. It is apparent
that none of the existing functionals can reproduce the universal
dependence in the low-density limit. This is expected, because the
functionals were not constructed to explore this regime. However, as
one approaches nuclear densities the discrepancies persist. Even the
SGII, SkO$'$, and SkM$^*$ functionals, which come close to the QMC
results around $k_F \sim 0.5 \, \text{fm}^{-1}$, have a stronger
density dependence than the microscopic results, and the Gogny D1N
functional, which was fit to neutron matter calculations, even differs
most from the microscopic $E_\text{pol}$ results.

Figure~\ref{fig:EDF} demonstrates that the polaron energy provides a
novel constraint for nuclear density functionals. To this end, we
construct a new density functional UNEDF1-pol, which we fit to the QMC
$E_\text{pol}$ results. Because the new exponent $\delta$ appears only
in the terms containing $\mathbf{s}$, this allows to fit the QMC
results without affecting the time-even part (i.e., the spin-symmetric
properties), for which we take the UNEDF1 functional~\cite{UNEDF1}.
Because the density dependence of the spin Skyrme parameters $C^s$
impacts the spin response of nuclei, we constrain the fit to reproduce
the sum of Landau parameters~\cite{Bender:2002} $G_0+G'_0=2.0$,
which also avoids possible spin instabilities in nuclear matter. This
value was chosen from microscopic calculations of asymmetric nuclear
matter~\cite{Zuo:2003}, because spin resonances are weak, making
it difficult to extract $G_0$ from experiment. We find
$C_0^{s,0}+C_1^{s,0} = 48 \pm 24 \, \text{MeV} \, \text{fm}^3$,
$C_0^{s,D}+C_1^{s,D} = 61 \pm 14 \, \text{MeV} \, \text{fm}^{3+\delta}$,
and $\delta = -0.288 \pm 0.026$. The large errors on the $C_0^s+C_1^s$
can be traced to a high correlation (0.994) with $\delta$, small
variations of which significantly change the polaron energy.

In summary, we compute the energy of the neutron polaron, generalizing the
polaron of condensed matter physics and ultracold atoms to strong interactions
with a significant effective range. We use QMC calculations to benchmark
non-perturbative contributions to the polaron energy, finding that for densities
$k_F \lesssim 0.5 \, \text{fm}^{-1}$ the polaron energy $E_\text{pol}$ and its
density dependence agree with results based on a dEFT and with microscopic
calculations using chiral EFT interactions at N$^3$LO that include contributions
beyond the effective range. Our results for the neutron polaron provide an anchor for the equation of state
of the neutron-star crust, in particular for the chemical potentials, which are also
important for thermal and transport properties (see, e.g., Refs~\cite{Gezerlis3,Aguilera}).

Finally, we show that current nuclear density functionals do not correctly
describe the neutron polaron, and construct a new functional UNEDF1-pol (based
on the UNEDF1~\cite{UNEDF1} functional) that satisfies the QMC constraints by
adjusting the time-odd components.  These results provide new constraints to
guide density functional theory to better describe polarized systems.

\begin{acknowledgments}
  We thank A.\ Bulgac, S.\ Gandolfi, and P.\ Massignan for discussions. This
  work was supported by the Helmholtz Alliance Program of the Helmholtz
  Association, contract HA216/EMMI ``Extremes of Density and Temperature: Cosmic
  Matter in the Laboratory'', the ERC Grant No.~307986 STRONGINT, the Natural
  Sciences and Engineering Research Council of Canada, the US DOE Grant
  No.~DE-FG02-00ER41132, the NSF Grants No.~PHY-0835543, PHY-1002478, and the
  Polish National Centre for Research and Development within the SARFEN--NUPNET
  Grant.  Computations were carried out at NERSC and at the J\"{u}lich
  Supercomputing Center. We thank the Institute for Nuclear Theory at the
  University of Washington for its hospitality and the US DOE for partial
  support.
\end{acknowledgments}

\end{document}